\documentclass[12pt]{article}

\newcommand{\be}{\begin{equation}}
\newcommand{\ee}{\end{equation}}
\newcommand{\al}{\alpha}
\newcommand{\ka}{\kappa}
\newcommand{\eps}{\varepsilon}
\begin{document}
\thispagestyle{empty}
\begin{flushright}
MZ-TH/99-52\\
November 1999
\end{flushright}
\vspace{0.5cm}
\begin{center}
{\Large\bf On the nonrelativistic dynamics 
of heavy particles near the 
production threshold}\\[1truecm]
{\large A.A.Pivovarov}\\[.7cm]
Institut f\"ur Physik, Johannes-Gutenberg-Universit\"at,\\
  Staudinger Weg 7, D-55099 Mainz, Germany\\[.2truecm]

and 

Institute for Nuclear Research of the\\
  Russian Academy of Sciences, Moscow 117312
\vspace{1truecm}
\end{center}

\begin{abstract}
A solution to the Schr{\"o}dinger equation
for the nonrelativistic
Green function
which is used for describing 
the heavy quark-antiquark pair production near the threshold in
$e^+e^-$ annihilation is presented.
A quick comparison with existing results is given.
A choice of the effective mass scale for the nonrelativistic
system
with Coulomb interaction is discussed.\\[4mm]
\end{abstract}

\newpage
The hadron production near heavy quark 
threshold 
will be thoroughly studied experimentally at future accelerators,
e.g.~\cite{exp}.
The dynamics of a slow moving
pair of the heavy quark and antiquark
near 
the production threshold is nonrelativistic to 
high accuracy that justifies the use 
of the nonrelativistic quantum mechanics as a proper 
theoretical framework for describing such a system \cite{CasLep}. 
Being much simpler than the comprehensive relativistic treatment 
this approach allows one to take into account 
exactly such essential features 
of the dynamics as Coulomb 
interaction \cite{VL,FK}.
The spectrum of hadronic states produced near 
the quark-antiquark threshold
is contained in the Green function 
$G(E)=(H-E)^{-1}$
of the effective nonrelativistic Hamiltonian $H$.
In the position space it reads
\be
G(E;{\bf r},{\bf r'})=\langle {\bf r}|(H-E)^{-1}| {\bf r'}\rangle
\label{position}
\ee
and satisfies the  Schr{\"o}dinger equation
\be
(H-E)G(E;{\bf r},{\bf r'})=\delta({\bf r}-{\bf r'})\, .
\label{Schr1}
\ee
The Hamiltonian $H$ 
is
represented in the following general form
$
H=H_0+\Delta H\, 
$
where the first term $H_0$ is  
the Coulomb Hamiltonian
\be
H_0=\frac{p^2}{m}-\frac{C_F \al_s}{r}
\ee
with $p^2=-\Delta^2$. 
The color factor for the fermion 
representation of the gauge group is $C_F=(N_c^2-1)/2N_c$,
$\al_s$ is a QCD coupling constant,
$m$ is a mass of the heavy quark. 
The normalization point for $\al_s$ will be fixed later.
The second term $\Delta H$
accounts for
relativistic and perturbative
strong interaction contributions
which are assumed to be small, i.e. 
they are treated as corrections to the Coulomb spectrum.
The term $\Delta H$
has an explicit form
\be
\Delta H=\Delta_k V
+\Delta_{pot}V
+\Delta_{NA}V+\Delta_{BF}V .
\label{Ham}
\ee
The quantity $\Delta_k V$ in eq.~(\ref{Ham})
is a relativistic kinetic energy 
correction
\be
\Delta_k V=-\frac{p^4}{4m^3}
\ee
the second term $\Delta_{pot}V$ is the 
strong interaction perturbative corrections to the
Coulomb potential \cite{Fish,Peter}, 
the third term $\Delta_{BF}V$
is a Breit-Fermi potential
with addition of the color factor $C_F$ \cite{AshBer,Landau}.
$\Delta_{NA}V=-C_AC_F\al_s^2/(2m r^2)$
is the non-Abelian potential
of quark-antiquark interaction \cite{Gup}, 
$C_A=N_c$ is a color factor for the gluons.
The corrections to the Coulomb Green function at the origin
due to terms $\Delta_kV$, $V_{NA}$ and $V_{BF}$ have first been
presented in \cite{Hoang,Mel}.
Numerous applications of these results 
and further references can be found in 
recent literature, e.g. \cite{hhh,MY,BSS1,diReg,PPres}.
The treatment of the problem in the present paper
is slightly different.
For the $s$-wave production of a quark-antiquark pair in 
the triplet spin state ($L=0$, $S=1$)
the Breit-Fermi potential reads \cite{AshBer,Landau} 
\[
\Delta_{BF}V = -\frac{C_F\al_s}{2m^2}\left(\frac{1}{r}p^2
+p^2\frac{1}{r}\right)+\frac{11\pi C_F\al_s}{3m^2}
\delta({\bf r})\, .
\]
The correction $\Delta H$ can be be rewritten as 
\[
\Delta H=\Delta_{pot}V
-\frac{1}{4m}H_0^2-\frac{3C_F\al_s}{4m}\left[H_0,\frac{1}{r}\right]_+
\]
\be
+\frac{\al_s}{m}\left(\frac{5C_F}{4}
+\frac{C_A}{2}\right)\left[H_0,i p_r\right]_-
-\frac{4\pi\al_s}{m^2}\left(\frac{C_F}{3}
+\frac{C_A}{2}\right)\delta({\bf r})
\label{red:corf}
\ee
where $ip_r=\partial_r+1/r$. 
Note that a similar form of the representation of the correction in terms
of powers of the leading order operator is usual for anharmonic
oscillator problems \cite{anh}.
The part of the potential proportional to 
the $\delta$-function $\delta({\bf r})$ is a separable potential.
The equation for the full
GF with such a potential can
be exactly solved for the quantity we need.
Noticing this fact one rewrites the Hamiltonian in the form 
\be
H=H_0+\Delta H=H_{ir}+\al_s V
\ee
with 
\be
V =-\frac{4\pi}{m^2}\left(\frac{C_F}{3}
+\frac{C_A}{2}\right)\delta({\bf r})=V_0\delta({\bf r}) \, .
\ee
Here $H_{ir}$ is an irreducible Hamiltonian.
The new representation for the GF reads
\be
G(E)=(H-E)^{-1}=(H_{ir}+\al_s V-E)^{-1}\, .
\ee
Introducing the GF $G_{ir}(E)$ of the irreducible Hamiltonian
\be
G_{ir}(E)=(H_{ir}-E)^{-1}
\ee
one obtains the following equation for the full Green function $G(E)$ 
(written in the operator form)
\be
G(E)=G_{ir}(E)-\al_s G_{ir}(E)VG(E) \, .
\label{ir:eq}
\ee
Eq.~(\ref{ir:eq}) is exactly solved for the position representation 
component $G(E;0,0)$ with the result
\be
\label{res:ir}
G(E;0,0)=\frac{G_{ir}(E;0,0)}{1+\al_s V_0 G_{ir}(E;0,0)}\, .
\ee
Eq.~(\ref{res:ir})
accounts for the $\delta({\bf r})$ part of the
correction to the irreducible Hamiltonian exactly.
It is analogous to the usual representation of the vacuum 
polarization function through the one particle irreducible block.
In this case the irreducible object $G_{ir}(E;0,0)$ is characterized 
by the absence of $\delta({\bf r})$ interaction in it.
The function $G_{ir}(E;0,0)$ can be found perturbatively
using the Coulomb solution as a leading order approximation.

The Green function $G(E)$ 
emerges as a nonrelativistic limit of
the relativistic scattering 
amplitude near the production threshold. 
The nonrelativistic Hamiltonian can be
constructed from the QCD Lagrangian. In the leading order of
nonrelativistic expansion there is an energy independent
factor
(matching coefficient) $C(\al_s,m)$
that allows one to map the quantum mechanical quantities 
onto the relativistic cross section near the threshold.
In higher orders of nonrelativistic expansion further terms
of the expansion of the current itself and new vertices 
of the effective Lagrangian are
generated 
that should be accounted for when the cross section
in QCD is calculated.
We do not discuss these terms
here because their contributions start at the higher orders 
in expansion parameters. Therefore the corresponding expressions 
should be taken only  
in the leading order of hard loop expansion.
The cross section 
of the heavy quark-antiquark pair
production near the threshold in $e^+e^-$ annihilation
contains a part with the nontrivial loop expansion.
It has the form
\be
R^{th}(s) \sim C(\bar\al_s,m){\rm Im}~G(E;0,0)\, ,
\quad s=(2m+E)^2 \, .
\label{cross}
\ee
Here $C(\bar\al_s,m)$ is the matching (hard or high energy)
coefficient,
$\bar\al_s$ is the coupling constant.
Generally, one can use the different
normalization points (or different subtraction procedures)
for $\bar\al_s$ and for the corresponding 
coupling constant $\al_s$ which enters the 
expression for the nonrelativistic Green function.
The vacuum polarization function near the threshold 
$C(\bar\al_s,m) G(E;0,0)$
requires subtraction.
This feature is familiar from 
the PT analysis of the vacuum polarization function in the 
full theory. 
To obtain a finite quantity one can
differentiate $C(\bar\al_s,m)G(E;0,0)$
with regard to $E$ (constructing the $D$-function) 
or take the discontinuity across the physical cut
(constructing the imaginary part or cross section
$R^{th}(s)$).
Both $D$-function and $R^{th}(s)$ are finite.
We write 
\[
G(E;0,0)=\frac{1}{\al_s V_0}-\frac{1}{\al_s V_0}
\frac{1}{1+\al_s V_0G_{ir}(E;0,0)}
\]
and obtain the following expression for the $D$-function
\be
\label{dfunk}
C(\bar\al_s,m)\frac{d}{dE}G(E;0,0)
=\frac{C(\bar\al_s,m)}{(1+\al_s V_0G_{ir}(E;0,0))^2}
\frac{d}{dE}G_{ir}(E;0,0)\, .
\ee
The imaginary part of $C(\bar\al_s,m)G(E;0,0)$ reads
\be
\label{im:res}
C(\bar\al_s,m){\rm Im}~G(E;0,0)
=
\frac{C(\bar\al_s,m){\rm Im}~G_{ir}(E;0,0)}
{(1+\al_s V_0 {\rm Re}~G_{ir}(E;0,0))^2
+(\al_s V_0 {\rm Im}~G_{ir}(E;0,0))^2}\, .
\ee
The quantities in eqs.~(\ref{dfunk})
and (\ref{im:res}) are finite.
The explicit expression for the coefficient 
$C(\bar\al_s,m)$ has been found
at the $\bar \al_s^2$ order within dimensional regularization \cite{coef}. 
It contains a singularity of the form
\be
\label{sing}
C_S(\bar\al_s,m)=1-C_F\bar\al_s^2 \left(\frac{C_F}{3}
+\frac{C_A}{2}\right)\frac{1}{\eps}
\ee
where only the singular part $C_S(\bar\al_s,m)$ of the coefficient 
$C(\bar\al_s,m)$ is written. This singularity cancels in 
eqs.~(\ref{dfunk}) and (\ref{im:res}).
We consider this cancellation 
(the renormalization procedure) for the case of $D$-function
only.
To fulfill the renormalization procedure 
we use 
the dimensionally regularized Green function in the expression 
for the correction 
proportional to $V_0$. 
It suffices to substitute the pure Coulomb Green function for 
this purpose.
The dimensionally regularized
Coulomb Green function at the origin
takes the explicit form \cite{diReg}
\be 
G_C^{DR}(\ka;0,0)={m\over 4\pi}\left\{-\ka 
+\frac{C_F \al_s m}{2}\left(\frac{1}{\eps}
+\ln\frac{\mu^2}{\ka^2}-
2\psi\left(1-\frac{C_F \al_s m}{2\ka}\right)\right)\right\}
\label{dimreg}
\ee
where $\ka^2=-mE$, $\psi(z)=\Gamma'(z)/\Gamma(z)$ is digamma function
and $\Gamma(z)$ is Euler's $\Gamma$-function.
One finds
\be
(1+\al_s V_0G_{ir}(E;0,0))^2\rightarrow
(1+\al_s V_0G_C^{DR}(E;0,0))^2
=Z^{-1}(1+\al_s V_0 G_C(E;0,0))^2
\ee
where 
\be
Z^{-1}=1+C_F\frac{\al_s^2}{4\pi} m^2V_0 \frac{1}{\eps}
=
1-C_F\al_s^2 \left(\frac{C_F}{3}
+\frac{C_A}{2}\right)\frac{1}{\eps}\, .
\ee
The finite (renormalized)
Coulomb Green function at the origin
has the form
\be 
G_C(\ka;0,0)={m\over 4\pi}\left\{-\ka 
+\frac{C_F \al_s m}{2}\left(\ln\frac{\mu^2}{\ka^2}-
2\psi\left(1-\frac{C_F \al_s m}{2\ka}\right)\right)\right\}\, .
\label{ren:dimreg}
\ee
The renormalization constant $Z$
cancels the divergence $C_S(\bar\al_s,m)$ of the high energy
coefficient $C(\bar\al_s,m)$ in a proper order of loop expansion.
One uses $\bar\al_s=\al_s+O(\al_s^2)$ as a formal PT relation
to achieve the cancellation.
The finite coefficient of order $\al_s^2$ (or ${\bar \al_s}^2$)
depends on particular ways of subtraction in 
$C(\bar\al_s,m)$ and $G_{ir}(E)$. 
If the subtraction procedures for 
$C(\bar\al_s,m)$ and $G_{ir}(E)$ are not properly coordinated
during the calculation
the finite coefficient is fixed by matching \cite{hoa:match}.
If the calculation for both $C(\bar\al_s,m)$ and $G_{ir}(E)$
has been done within one and the same subtraction scheme this matching
is automatic, e.g. \cite{barH}. 
Therefore after the standard renormalization,
eq.~(\ref{res:ir})
gives the representation of $G(E;0,0)$ as 
a Dyson sum of irreducible terms.
The spectrum of the full system is determined by 
the equation 
\be
\label{spec:eq}
G_{ir}(E;0,0)^{-1}+\al_s V_0 = 0
\ee
where $G_{ir}(E;0,0)$ is constructed perturbatively.
Eq.~(\ref{spec:eq})
can be solved exactly or perturbatively.
The isolated roots of eq.~(\ref{spec:eq})
give the discrete spectrum of the system.
There is also a continuous spectrum given by the discontinuity of 
$G_{ir}(E;0,0)$ across the cut at positive values of $E$.
For the continuous spectrum one can use 
the Coulomb GF in the denominator of 
eq.~(\ref{im:res}). 

The problem of calculating the near-threshold cross section
reduces to the construction 
of the spectrum of the irreducible Hamiltonian $H_{ir}$
\be
H_{ir}=H_0+\Delta V_{pot}
-\frac{1}{4m}H_0^2-\frac{3C_F\al_s}{4m}\left[H_0,\frac{1}{r}\right]_+
+\frac{\al_s}{m}\left(\frac{5C_F}{4}
+\frac{C_A}{2}\right)\left[H_0,i p_r\right]_- \, .
\label{irrham}
\ee
The spectrum is found within PT.
The leading order spectrum is given by 
the renormalized pure Coulomb solution
eq.~(\ref{ren:dimreg}).
The term $\Delta_{pot}V$ represents
the first and second order perturbative QCD corrections to the
Coulomb
potential 
which were studied.
The correction due to the first iteration
of the $\Delta_{pot}V$ term
has been found in ref.~\cite{KPP}
where the simple and efficient framework for computing
the iterations of higher orders was formulated.
The explicit 
formulas for these corrections can be found in \cite{PPres}.
For the kinetic $H_0^2$ term one finds
\be
\rho_k(E)=\sum_{E'} \delta(E'-\frac{{E'}^2}{4m}-E)\rho(E')
\ee
where $\rho(E)$ is a density of Coulomb states with the energy $E$
\be
\rho(E)=\frac{1}{\pi}{\rm Im}~G_C(E;0,0)\, .
\ee
The sum is over the whole spectrum.
For the discrete levels it gives
\be
\rho_k(E)|_{disc}=\sum_{n} \delta(E_n-\frac{E_n^2}{4m}-E)|\psi_n(0)|^2
\ee
with $\psi_n$ being a bound state with the energy level $E_n$.
The position of the pole is now at $E_n^{pole}=E_n-E_n^2/4m$.
For the continuous spectrum the correction reads
\be
\rho_k(E)|_{cont}=\int_0^\infty dE' 
\delta(E'-\frac{{E'}^2}{4m}-E)\rho(E')
=\frac{1}{1-\bar E/2m}\rho(\bar E)
\ee
with 
\be
\label{ki:cor}
E-\bar E + \frac{{\bar E}^2}{4m}=0\, .
\ee
Eq.~(\ref{ki:cor}) can be solved perturbatively
for small $E$. One obtains
\be
\bar E=E+ \frac{{E}^2}{4m} + O(E^3)
\ee
which is the substitution used in refs.~\cite{Hoang,Mel}.
Note that there is no correction to wave functions
because $H_0^2$ has no non-diagonal matrix elements
between the Coulomb states.
The last term in eq.~(\ref{irrham})
gives no correction
to the spectrum.
The term with the anticommutator 
$[H_0,r^{-1}]_+$ in 
eq.~(\ref{irrham}) gives the correction
to the spectrum which was presented earlier as an energy dependent 
shift of the coupling constant.
This correction can be written in the
form
\be
G_C(E)+\Delta_a G_C(E)
=
G_C\left(E;\al_s \rightarrow \al_s\left(1+\frac{3E}{2m}\right)\right)
\label{alch}
\ee
which is valid at small $E$ and was presented in \cite{Hoang,Mel}.
Note, however, that while the $\delta$-function part 
of the correction to the Hamiltonian can be considered unique
because it represents a reducible vertex, the irreducible potential 
$H_{ir}$ can be chosen in different forms.
Indeed, one can rewrite the sum of kinetic and anticommutator
terms of $H_{ir}$
in the form
\be
-\frac{1}{4m}H_0^2-\frac{3C_F\al_s}{4m}\left[H_0,\frac{1}{r}\right]_+
=\frac{5}{4m}H_0^2-\frac{3}{4m^2}\left[H_0,p^2\right]_+
\ee
which is a reshuffling of contributions within the irreducible correction.
While the kinetic term $H_0^2$
has only its coefficient changed as a result of
such a reshuffling, the new
anticommutator term
leads to a modification of the leading order GF of the following form
\[
\frac{1}{H_0-E}+\frac{3}{4m^2}
\left(p^2\frac{1}{H_0-E}+\frac{1}{H_0-E}p^2\right)
+\frac{1}{H_0-E}E\frac{3p^2}{2m^2}\frac{1}{H_0-E}
\]
\be
=
\frac{3}{4m^2}
\left(p^2\frac{1}{H_0-E}+\frac{1}{H_0-E}p^2\right)
+
\left(\frac{p^2}{m}\left(1-\frac{3E}{2m}\right)
-\frac{C_F\al_s}{r}-E\right)^{-1}\, .
\label{newdec}
\ee
The first term of this equation does not affect the structure of the
spectrum.
The last term in eq.~(\ref{newdec})
can be interpreted as a correction to the mass.
To a certain degree this is equivalent to the previous case
(the correction to the coupling $\al_s$ in eq.~(\ref{alch}))
because the genuine parameter of the Coulomb
problem is the Bohr radius or momentum, $p_B=\al_s m$
(or $C_F\al_s m/2$).
One can see that both changes 
$\al_s \rightarrow \al_s\left(1+3E/2m\right)$ 
and $m\rightarrow m/\left(1-3E/2m\right)$ lead to the same 
result for $p_B$ within the accuracy of the approximation, 
i.e. up to higher order terms in $E$
\be
p_B\rightarrow p_B\left(1+\frac{3E}{2m}\right)
=\frac{p_B}{1-\frac{3E}{2m}}+O(E^2)\, .
\ee
One can check that the total correction to the discrete energy levels,
for instance, is the same.
The old solution eq.~(\ref{alch}) gives
\be
m\Delta E_n=-\frac{1}{4}E_n^2+3E_n^2=\frac{11}{4}E_n^2
\ee
while the new decomposition eq.~(\ref{newdec})
results in 
\be
m\Delta E_n=\frac{5}{4}E_n^2+\frac{3}{2}E_n^2=\frac{11}{4}E_n^2\, .
\ee
This coincidence is valid only parametrically in the region 
of applicability of perturbation theory in 
the expansion parameters $E/m$ and/or $\al_s$.
It can result in numerical difference 
when extrapolated to larger energies
in the continuous spectrum.
The different forms of the decomposition 
may lead to different numerical predictions within numerical evaluation 
\cite{Hoang,Mel,hhh,ttee,KuhTeu}.

The expression given in eq.~(\ref{res:ir})
has a standard structure of the polarization function. 
The solution $E_f$ to
eq.~(\ref{spec:eq}) (one or the first of the poles 
of the generalized ``propagator''
of the full system with the Hamiltonian $H$) is an important 
dimensional 
parameter for the system. 
For the observables which are saturated with the contributions of the
discrete spectrum the quantity $E_f$ can serve
as a natural mass parameter. 
In this case the position of a pole 
(the numerical value of some solution $E_f$ to
eq.~(\ref{spec:eq})) can be chosen as a scale for the system instead
of the heavy quark pole mass.
However, for the observables which are 
saturated with the contributions of the
continuous spectrum or have a considerable admixture
of such contributions, the natural mass scale 
is not necessarily related to $E_f$
and is determined by other properties of 
the full interaction. 

To conclude, we have presented a solution to the
Schr{\"o}dinger equation for the nonrelativistic Green function.
The solution has the form of a resumed geometric 
series (Dyson resummation) of irreducible blocks
which is 
usual for quantum field theory. The property of irreducibility is defined 
with respect to the $\delta$-function part of the interaction
potential. 
Different decompositions of the irreducible Hamiltonian $H_{ir}$
for the treatment within PT are considered.
A new form 
of the first order correction to 
the spectrum of irreducible Hamiltonian $H_{ir}$
is given.

\vspace{5mm}
\noindent
{\large \bf Acknowledgments}\\[2mm]
This work is partially supported
by Volkswagen Foundation under contract
No.~I/73611 and Russian Fund for Basic Research under contract
Nos.~97-02-17065 and 99-01-00091. 
A.A.Pivovarov is
Alexander von Humboldt fellow.

\end{document}